\documentclass{iaus}
\usepackage{graphicx}

\title[Mass dependent Evolution of Field Early-Type Galaxies Since $z=1$] 
{Mass dependent Evolution of \\ Field Early-Type Galaxies Since $\mathbf{z=1}$}

\author[A. Fritz, I. J{\o}rgensen \& R.\ P.\ Schiavon]   
{Alexander Fritz, Inger J{\o}rgensen, \and  Ricardo P.~Schiavon}

\affiliation{Gemini Observatory, 670 N.\ A'ohoku Place,
Hilo, HI 96720, USA \\
e-mail: {\tt afritz@gemini.edu}}

\pubyear{2009}
\volume{262}  
\pagerange{1--2}
\jname{Stellar Populations - Planning for the Next Decade}
\editors{G.R. Bruzual, \& S. Charlot, eds.}
\begin{document}

\maketitle

\begin{abstract}
We present the Fundamental Plane (FP) of field early-type galaxies at
$0.5<z<1.0$. Our project is a continuation of our efforts to understand the formation and
evolution of early-type galaxies in different environments.
The target galaxies were selected from the comprehensive and
homogeneous data set of the Gemini/HST Galaxy Cluster Project.
The distant
field early-type galaxies follow a steeper FP relation compared to the local FP.
The change in the slope of the FP can be interpreted as a mass-dependent
evolution. Similar results have been found for cluster early-type galaxies in
high redshift galaxy clusters at $0.8<z<1$. Therefore, the slope change of the
FP appears to be independent of the environment of the galaxies. 

\end{abstract}

\firstsection 

\section{Introduction}

Early-type galaxies obey a tight linear relation in 3-D
log-space, the Fundamental Plane (FP; Djorgovski \& Davis 1987;
Dressler et al.\ 1987), defined by their 
size $r_{{\rm e}}$, average surface brightness 
$\langle I_{{\rm e}}\rangle$ and stellar velocity
dispersion ($\sigma$). Via a study of the FP constraints on the
formation epoch and evolution of early-type galaxies are possible.

Over the past twenty years a significant effort has been devoted to
understand the physical processes involved in this empiric
relationship in the nearby universe. 
Previous works of field early-type galaxies at higher redshift up to
$z\sim1.2$ found evidence for a mass dependent evolution
(e.g., Treu et al.  2005; di Serego Alighieri et al. 2006).
Recently, Fritz et al. (2009a) detected recent star formation
in less-massive field early-type galaxies up to $z\sim0.8$.
This rejuvenation accounts for 5-10\% in the total stellar mass
budget of these galaxies, possible triggered through AGN feedback.

\section{Observational Data}

Target galaxies were selected from the Gemini/HST Galaxy Cluster Project.
This project is an extensive observational campaign to
investigate the formation and evolution of distant galaxies in rich galaxy
clusters from $z=1$ to the present-day (J\o rgensen et al.\ 2005).
For 15 massive, $X$-ray luminous galaxy clu\-sters 
($L_X\,{\rm (0.1-2.4\,keV)}\le 2\times10^{44}$ erg s$^{-1}$)
from $z=0.1$ to 1, intermediate-resolu\-tion Gemini/GMOS spectroscopy
and HST/ACS or WFPC2 ima\-ging were obtained. The MOS masks included
primarily expected cluster members, while vacant space was filled with slits
for objects that have similar brightnesses but slightly bluer or redder
colours than expected for early-type cluster members. Out of three cluster
fields, RXJ0152.7-1357, RXJ1226.9+3332 and RXJ1415.1+3612,
we have extracted 20 galaxies with early-type morphology and $0.5<z<1$
($\langle z\rangle=0.74$) which are used in the following. The high
$S/N$ and intermediate resolution galaxy spectra
($\langle S/N\rangle \sim 25\,\AA^{-1}_{\rm rest}$)
allows us to study in detail the internal kinematics and stellar
pop\-ulations of the galaxies. Using deep HST/ACS imaging, we fit model
profiles to the galaxies to derive the effective parameters
(Chiboucas et al. 2009).

\section{The FP of Field Early-Type Galaxies up to $z=1$}

\begin{figure}[t]
\begin{center}
 \includegraphics[width=4.8cm, angle=-90]{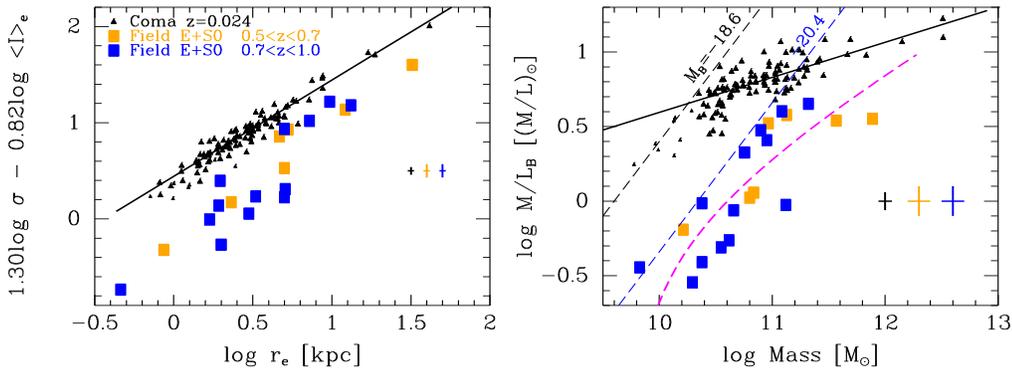} 
 \vspace*{0.2cm}
 \caption{Fundamental Plane (FP) of field early-type galaxies in two redshift
bins $0.5<z<0.7$ (orange) and $0.7<z<1.0$ (blue). 
Left: Edge-on projection of FP. The Coma cluster is used as a local reference
(black triangles, solid line, J{\o}rgensen 1999). 
Right: $M/L-M$ plane. Dashed
lines indicate the magnitude limits of Coma (black) and the distant field
galaxies (large symbols). SSP model predictions are indicated as the dashed
magenta line (Thomas et. al. 2005).}
   \label{fig1}
\end{center}
\end{figure}

Fig.~\ref{fig1} shows the FP for field early-type galaxies (squares)
in rest-frame Johnson $B$-band, compared to 116 early-type
galaxies in the Coma cluster (triangles). The slope of the FP for
field galaxies at high redshift is steeper than that for their local
counterparts. We interpret the slope difference between the distant and 
local relation as a mass-dependent evolution. Less-massive field galaxies
($M\le 8\times10^{10}M_{\odot}$) evolve faster than their more massive
counterparts. The $M/L$ evolution with $z$ is under investigation.
Similar results have been found for galaxy clusters at $0.8<z<1.0$
(J\o rgensen et al.\ 2006, 2007; Fritz et al.\ 2009b). This suggest that
environmental effects play only a minor role on the evolution of
early-type galaxies. A combined analysis of the FP and absorption
line strengths for the field galaxies will be presented in
Fritz et al.\ (2009c, in prep).

\vspace*{0.2cm}

AF and IJ acknowledge support from grant HST-GO-10826.01 from STScI. 
Based on observations obtained at the Gemini Observatory, 
operated by AURA, Inc., under cooperative agreement with the NSF, 
on behalf the NSF, STFC (UK), NRC (Canada), CONICYT (Chile),
ARC (Australia), CNPq (Brazil), and SECYT (Argentina). 


\vspace*{-0.4cm}

\begin{thebibliography}{}

\bibitem[Chiboucas et al.\(2009)]{CJ09}
Chiboucas, K., Barr, J., Flint, K. et al. 2009, ApJS, 184, 271

\bibitem[\protect\citeauthoryear{di Serego Alighieri et al.}{2006}]{SLJ06}
di Serego Alighieri, S., Lanzoni, B., J{\o}rgensen, I. 2006, ApJ, 652, L145

\bibitem[\protect\citeauthoryear{Djorgovski \& Davis}{1987}]{DD87}
Djorgovski, S., \& Davis, M. 1987, ApJ, 313, 59

\bibitem[\protect\citeauthoryear{Dressler et al.}{1987}]{Dre:87}
Dressler, A., et al. 1987, ApJ, 313, 42 

\bibitem[\protect\citeauthoryear{Fritz et al.}{2009}]{FBZ09}
Fritz, A., B\"ohm, A., \& Ziegler, B.~L. 2009a, MNRAS, 393, 1467

\bibitem[\protect\citeauthoryear{Fritz et al.}{2009}]{FJSC09}
Fritz, A., J{\o}rgensen, I., Schiavon, R.~P., \& Chiboucas, K.
2009b, AN, 330, 931

\bibitem[\protect\citeauthoryear{J{\o}rgensen}{1999}]{J99}
J{\o}rgensen, I. 1999, MNRAS, 306, 607

\bibitem[\protect\citeauthoryear{J{\o}rgensen et al.}{2005}]{JB05}
J{\o}rgensen, I., Bergmann, M., Davies, R., et al. 2005, AJ, 129, 1249

\bibitem[\protect\citeauthoryear{J{\o}rgensen et al.}{2006}]{JCFBBD06}
J{\o}rgensen, I., et al. 2006, ApJ, 639, L9

\bibitem[\protect\citeauthoryear{J{\o}rgensen et al.}{2007}]{JCB07}
J{\o}rgensen, I., et al. 2007, ApJ, 654, L179

\bibitem[\protect\citeauthoryear{Thomas et al.}{2005}]{TMB05}
Thomas, D., et. al. 2005, ApJ, 621, 673

\bibitem[\protect\citeauthoryear{Treu et al.}{2005}]{T05}
Treu, T., et al. 2005, ApJ, 633, 174

\end{thebibliography}
\end{document}